\documentclass[twocolumn,nofootinbib,superscriptaddress,preprintnumbers]{revtex4-1}

\pdfoutput=1

\let\ifpdf\relax
\usepackage{graphicx}

\usepackage{slashed}
\usepackage{color}
\def\be{\begin{equation}}
\def\ee{\end{equation}}

\newcommand{\bea}{\begin{eqnarray}}
\newcommand{\eea}{\end{eqnarray}}

\usepackage{bbm,bm,amsmath,amssymb}

\usepackage{hyperref}
\usepackage{comment}

\newcommand{\Mpl}{M_\text{Pl}}

\begin{document}

\preprint{CTPU-PTC-24-34,
KYUSHU-HET-302}

\title{Inflationary constraints on the moduli-dependent species scale\\ in modular invariant theories}

\author{Shuntaro Aoki}
\email{shuntaro1230@gmail.com}
\affiliation{
Center for Theoretical Physics of the Universe, Institute for Basic Science (IBS), Daejeon
34051, Korea}
\author{Hajime Otsuka} 
\email{otsuka.hajime@phys.kyushu-u.ac.jp}
\affiliation{
Department of Physics, Kyushu University, 744 Motooka, Nishi-ku, Fukuoka 819-0395, Japan}

 \begin{abstract}
\noindent
We demonstrate that a broad class of modular inflation models predicts the emergence of new physics within an energy range of approximately \( 10^{15} \, \mathrm{GeV} \) to \( 10^{17} \, \mathrm{GeV} \). This prediction arises by comparing the moduli-dependent species scale with observational constraints on inflation. Specifically, we illustrate this within the context of \( SL(2, \mathbb{Z}) \)-modular inflation models by re-expressing inflationary observables in terms of the species scale. We further discuss the implications of this approach for generic Calabi-Yau threefolds.

  \end{abstract}

\maketitle

\vspace{1cm}

\section{Introduction}
A large \( N \) number of light degrees of freedom in a quantum theory of gravity modifies the scale at which quantum effects of gravity become relevant. Such an effective ultraviolet cutoff in theories of quantum gravity is called the species scale~\cite{Dvali:2007hz, Dvali:2007wp, Dvali:2008ec}, which in four dimensions is given by
\begin{align}
    \Lambda_{\rm sp} = \frac{M_{\rm Pl}}{N^{1/2}}.
\label{eq:L_sp}
\end{align}
Since a large number of species leads to a decrease in \( \Lambda_{\rm sp} \), the quantum effects of gravity become relevant below the Planck scale.

The species scale depends on the light modes \( \tau \), which in string compactifications correspond to moduli fields exhibiting geometric symmetries of compact extra-dimensional spaces.\footnote{We have focused on the geometric moduli throughout this paper.} 
It was shown in Refs.~\cite{vandeHeisteeg:2022btw, vandeHeisteeg:2023ubh, Castellano:2023aum, vandeHeisteeg:2023dlw} that the moduli-dependent species scale is constrained to be an automorphic form of the duality symmetries, including the modular symmetries of the theory, as calculated in Refs.~\cite{Green:1999pu, Green:1999pv, Green:2005ba, Green:2010kv, Green:2010wi}. 
In type II compactifications on Calabi-Yau (CY) threefolds, the species scale can be identified with the genus-one topological free energy \( F_1 \), as proposed in Ref.~\cite{vandeHeisteeg:2022btw}, i.e., \( N \simeq F_1 \), which is described by a specific modular function, as explicitly demonstrated in the Enriques CY \( (K3 \times T^2)/\mathbb{Z}_2 \) \cite{Ferrara:1995yx}.

The decrease of the species scale impacts particle phenomenology and cosmology. 
Indeed, the decay rate of the tower of states \( |\partial_\phi \Lambda_{\rm sp} / \Lambda_{\rm sp}| \) 
is found to have a lower bound~\cite{vandeHeisteeg:2023ubh, Calderon-Infante:2023ler, Castellano:2023stg, Castellano:2023jjt} and an upper bound of order \( {\cal O}(1) \)~\cite{vandeHeisteeg:2023ubh, Calderon-Infante:2023ler, vandeHeisteeg:2023dlw, Lust:2023zql}, which permits only a finite range for the light mode in gravitational effective field theories. 
By imposing this finite range on the inflaton field during an accelerated expansion of the universe, it has been shown that the inflaton field range is bounded by the tensor-to-scalar ratio~\cite{Cribiori:2023sch, Scalisi:2019gfv, vandeHeisteeg:2023uxj, Scalisi:2024jhq}. 
Furthermore, the coefficient of the \( {\cal R}^2 \) term becomes a duality-invariant form, and the Starobinsky inflation model has been revisited from the perspective of string theory~\cite{Lust:2023zql}.

In this paper, we focus on the \( SL(2, \mathbb{Z}) \) modular symmetry, which appears in four-dimensional effective field theories (EFT) on toroidal compact spaces and in the asymptotic limits of CY moduli spaces \cite{Ishiguro:2024xph}. 
By imposing \( SL(2, \mathbb{Z}) \) modular invariance on the theory, one can achieve a unique pattern of flavor structure~\cite{Feruglio:2017spp} and a successful inflation mechanism~\cite{Kobayashi:2016mzg, Schimmrigk:2016bde, Abe:2023ylh, Ding:2024neh, King:2024ssx, Casas:2024jbw, Kallosh:2024ymt}, where the scalar potential is expected to be stable against higher-order corrections, in contrast to the Starobinsky model.
Remarkably, several cosmological observables, such as the tensor-to-scalar ratio, are 
determined by the modular-invariant species scale~\cite{Casas:2024jbw}, indicating that recent cosmological observations 
place a bound on the species scale itself as well as its decay rate. 
Unlike the decay rate of the species, a phenomenologically viable range of the species scale has not yet been fully explored. 
The purpose of this {\it Letter} is to evaluate the species scale directly in modular-invariant theories by utilizing 
recent cosmological observations. 
We find that the species scale is restricted to the range \( 10^{15} \, \text{GeV} \lesssim \Lambda_{\rm sp} \lesssim 10^{17} \, \text{GeV} \) when the spectral index and tensor-to-scalar ratio are required to match observational data.

\section{Moduli-dependent species scale}
The cutoff scale of gravitational EFT is defined as the scale at which gravitational corrections to Einstein gravity become relevant. 
The effective action can be written as
\begin{align}
    S_{\rm EFT} \supset \int d^4 x \sqrt{-g} \frac{M_{\rm Pl}^2}{2}\left( {\cal R} + \frac{{\cal O}_n({\cal R})}{\Lambda_{\rm sp}^{n-2}}\right),
\label{eq:EFT}
\end{align}
up to the matter sector, including the Standard Model. Here, \( {\cal O}_n({\cal R}) \) denotes higher-curvature corrections with mass dimension \( n > 2 \). 
When there exists a large number of species, the cutoff scale of gravitational EFT is described by the species scale \eqref{eq:L_sp} rather than the Planck scale. 
According to the Distance Conjecture~\cite{Ooguri:2006in}, such a large number of light species is expected to appear at the boundaries of moduli space. The definition of the species scale in the context of perturbative analysis of gravitational EFT is known to be consistent with that in Black Hole (BH) analysis~\cite{vandeHeisteeg:2022btw,Calderon-Infante:2023uhz}, where the area of BH horizons is modified by the presence of higher-derivative corrections induced by a large number of light species~\cite{Sen:2005wa}.

In string theory, following the emergent string conjecture~\cite{Lee:2019wij}, there are only two types of towers determining the species scale at the boundaries in the moduli space: a Kaluza-Klein tower or the states in a string. In the case of heterotic string theory, the species scale is determined by a tower of string states. 
On the other hand, in four-dimensional \( {\cal N}=2 \) effective theories, the species scale appearing in the one-loop \( {\cal R}^2 \) operator (\( n = 4 \) in \eqref{eq:EFT}) is determined by the A-model topological free energy at genus 1, \( F_1 \), up to an additive constant~\cite{vandeHeisteeg:2022btw}.

In CY compactifications of type IIA string theory, $F_1$ is described by vector multiplet moduli of ${\cal N}=2$ effective theory \cite{Bershadsky:1993ta}. 
For instance, in the Enriques CY $(K3 \times T^2)/\mathbb{Z}_2$, the number of species~$N$ was found to be \cite{Klemm:2005pd,Grimm:2007tm}:
\begin{align}
    N \simeq -6 \ln [2\tau_2 |\eta(\tau)|^4] + N_0,
\label{eq:N_enriques}
\end{align}
with $\tau_2 \equiv {\rm Im}\tau$, where $\eta(\tau)$ is the Dedekind eta function of the modulus $\tau$ and  $N_0$ is determined by the structure of $K3$ surface. In the large moduli regime $\tau_2 \rightarrow \infty$ corresponding to an emergent string limit, the number of species is described by\footnote{Note that the moduli-independent term in Eq. \eqref{eq:N_enriques} is irrelevant to the following discussion of inflation as long as $N_0\lesssim 100$ \cite{Casas:2024jbw}.}
\begin{align}
    N \simeq 2\pi \tau_2.
\label{N_sp}
\end{align}
In a generic CY case, there exist multiple moduli fields labeled by $\tau_2^i$. When we take the large field limit for the moduli fields under a parametrization $\tau_2^i = s \hat{\tau}_2^i$, following Ref.~\cite{vandeHeisteeg:2023uxj}, its asymptotic behavior is described as \cite{Bershadsky:1993ta}
\begin{align}
    F_1 = \frac{2\pi}{12} c_2^{({\rm CY})} s  +{\cal O}(\ln s),
\label{eq:F1_CY}
\end{align}
with $c_2^{({\rm CY})}=\sum_i c_2^i \hat{\tau}_2^i$, 
where $c_2^i$ denotes components of second Chern class of CY threefolds. 
It corresponds to the moduli-dependent species scale:
\begin{align}
    N \simeq \frac{2\pi}{12} c_2^{({\rm CY})} s.
\label{eq:N_sp_general}
\end{align}
Note that the modular-invariant
completion of Eq.~\eqref{eq:F1_CY} was proposed in Refs. \cite{LopesCardoso:1999cv,Cribiori:2023sch} as in the similar form of Eq.~\eqref{eq:N_enriques}. 
In the following analysis, Eqs.~\eqref{N_sp} and~\eqref{eq:N_sp_general} will be utilized in evaluating constraints on the species scale from inflationary prediction.

\section{\( SL(2, \mathbb{Z}) \)-modular inflaton models}
\label{sec:setup}

It was known that the low-energy effective action of string theory equips the modular symmetry associated with geometric symmetries of extra-dimensional spaces such as toroidal orbifolds \cite{Ferrara:1989qb,Lerche:1989cs,Lauer:1989ax,Lauer:1990tm} and Calabi-Yau manifolds \cite{Strominger:1990pd,Candelas:1990pi,Ishiguro:2021ccl}.

Let us consider on a modulus field determining the species scale as in Eq.~\eqref{eq:N_sp_general}. As an effective action of moduli fields, we consider the following Lagrangian for a modulus field $\tau=\tau_1+i\tau_2$:
\begin{align}
\frac{\mathcal{L}(\tau, \bar{\tau})}{\sqrt{-g}}=\frac{\Mpl^2}{2}R-\frac{\Mpl^2\partial \tau \partial \bar{\tau}}{2a^2\tau_2^2} -V(\tau, \bar{\tau}),    
\end{align}
where the scalar kinetic term possesses $S L(2, \mathbb{R})$ symmetry 
\begin{align}
\tau \rightarrow\frac{p \tau+q}{r \tau+s}, \quad \text { where }  \quad p s-r q=1,    
\end{align}
with $\{p,q,r,s\} \in \mathbb{R}$, which is broken down to $S L(2, \mathbb{Z})$ by the potential. 
Such \( SL(2, \mathbb{Z}) \)-invariant scalar potentials have been discussed in Refs.~\cite{Casas:2024jbw} (referred to as the CI model) and~\cite{Kallosh:2024ymt,Kallosh:2024pat} (referred to as the KL models). As pointed out in Ref.~\cite{Kallosh:2024ymt}, one can classify the potential by \( SL(2, \mathbb{Z}) \)-invariants:
\begin{align}
&L_j(\tau, \bar{\tau}) \equiv \frac{1}{4 \pi} \ln \left(\left|j(\tau)\right|^2+j^2(i)\right),\\
&L_\eta(\tau, \bar{\tau}) \equiv-\frac{3}{\pi} \ln \left(\tau_2\left|\eta(\tau)\right|^4\right),\\
&L_{G_2}(\tau, \bar{\tau}) \equiv\frac{3^2}{\pi^2}\left|\tau_2 \tilde{G}_2\right|^2,
\end{align}
where \( j(\tau) = 12^3 J(\tau) \), with \( J(\tau) \) denoting Felix Klein's absolute invariant such that \( J(i) = 1 \). Here, \( \tilde{G}_2 \equiv -4\pi i \, \partial_\tau \ln \eta(\tau) - \pi / \tau_2 \) is referred to as an almost holomorphic modular form of weight \( 2 \). Based on these quantities, models discussed in Refs.~\cite{Casas:2024jbw,Kallosh:2024ymt,Kallosh:2024pat} can be summarized as
\begin{align}
V(\tau, \bar{\tau})=V_0\times \left\{\begin{array}{l}
\left(\frac{L_{G_2}(\tau, \bar{\tau})}{\left(2 \pi L_\eta(\tau, \bar{\tau})+\tilde{N}_0\right)^2}\right)^n, \quad {\rm{CI-}model}\\
\left(\frac{I(\tau, \bar{\tau})-1}{I(\tau, \bar{\tau})+1}\right)^n, \quad {\rm{KL-T-}model}\\
\left(1-I^{-1}(\tau, \bar{\tau})\right)^n, \quad {\rm{KL-E-}model}
\end{array}\right.
,
\label{Models}  
\end{align}
where $V_0$ and $\tilde{N}_0$ are moduli-independent constant. 
The $I$-function in KL models can be either \( j \)-invariant: $I(\tau,\bar{\tau})=4\pi L_j(\tau, \bar{\tau})/\ln j^2(i)$ or \( \eta \)-invariant: $I(\tau,\bar{\tau})=-\pi L_\eta(\tau, \bar{\tau})/(3\ln \eta^4(i))$. 

Inflation occurs in a region \( \tau_1 = 0 \) and \( \tau_2 \gg 1 \).\footnote{We do not discuss the dynamics of $\tau_1$ in this paper. However, Ref.~\cite{Carrasco:2025rud} explores possible mechanisms to make the $\tau_1$ direction heavy without affecting the inflationary dynamics or violating the \( SL(2, \mathbb{Z}) \) symmetry.} For any case of Eq.~\eqref{Models}, the scalar potential for \( \tau_1 = 0 \) and \( \tau_2 \gg 1 \) can be expanded as
\begin{align}
V(\tau, \bar{\tau})\underset{\tau_2 \gg 1}{\simeq}  V_0\left(1-\frac{c_n}{\tau_2}+\cdots\right),
\end{align}
where 
\begin{align}
c_n=\left\{\begin{array}{l}
\frac{6n}{\pi}, \quad {\rm{CI-}model}\\
\frac{2n}{c}, \quad {\rm{KL-T-}model}\\
\frac{n}{c}, \quad {\rm{KL-E-}model}
\end{array}\right.    
.
\end{align}
A real parameter \( c \) is related to the normalization of the \( I \)-function: \( c = 4 \pi / \ln |j(i)|^2 \simeq 0.84 \) for the \( j \)-invariant, and \( c = -\pi/(3\ln \eta^4(i))\simeq0.99 \) for the \( \eta \)-invariant. For the CI model, we redefine \( V_0 \) by absorbing some numerical factors. In terms of the canonically normalized inflaton \( \tau_2 \equiv e^{a\phi / \Mpl} \) and by truncating sub-leading terms, the inflaton potential can be written as
\begin{align}
V_{\rm{inf}}(\phi)\simeq V_0 \left(1-c_n e^{-a\phi/\Mpl}\right). \label{V_inf}   
\end{align}
Note that such an inflaton potential with $a=\sqrt{2/(3\alpha)}$ and $c_n=1$ is related to $\alpha$-attractor models as discussed in Ref. \cite{Kallosh:2024ymt}.

Before closing this section, we provide evidence that the moduli potential considered in this paper may arise from string theory. It is known that the \( SL(2, \mathbb{Z}) \)-invariant function \(L_\eta(\tau, \bar{\tau})\) appears in threshold corrections to gauge coupling constants~\cite{Dixon:1990pc}, as well as in the one-loop graviton scattering amplitude in eight-dimensional supergravity~\cite{Green:2010wi}. A more general \( SL(2, \mathbb{Z}) \)-invariant function without singularities in the fundamental domain of $\tau$ was proposed in Ref.~\cite{Cvetic:1991qm} using the modular \(j\)-function. Although several additional moduli typically appear in string compactifications, we leave a comprehensive analysis of the species scale and moduli stabilization to future work.

\section{Determination of the species scale from inflationary prediction}\label{sec:inflation}
It is straightforward to compute the slow-roll parameters and inflationary observables~\cite{Casas:2024jbw, Kallosh:2024ymt}. Here, we re-express them in terms of the species scale defined in Eq.~\eqref{eq:L_sp} and determine it based on observations.

For large field regime $\phi/\Mpl\gg 1$, the slow-roll parameters $\epsilon$ and $\eta$ can be evaluated as
\begin{align}
&\epsilon\equiv \frac{M_{\mathrm{Pl}}^2}{2}\left(\frac{V^{\prime}}{V}\right)^2 \simeq \frac{ a^2 c_n^2 e^{-2 a \phi/\Mpl }}{2},\\
&\eta\equiv\frac{M_{\mathrm{Pl}}^2 V^{\prime \prime}}{V} \simeq-a^2 c_n e^{-a \phi/\Mpl},
\end{align}
where the prime denotes the derivative with respect to~$\phi$. Then, from the definition of the number of e-folds,
\begin{align}
\mathcal{N}= \frac{1}{\Mpl}\int_{\phi_e}^{\phi_*} \frac{1}{\sqrt{2\epsilon}} \, \mathrm{d} \phi \simeq \frac{e^{a \phi_*/\Mpl}-e^{a \phi_e/\Mpl}}{a^2 c_n},\label{N_e}
\end{align}
one can estimate the field value \( \phi_* \) at the time when the cosmic microwave background (CMB) scale exits the horizon, corresponding to \( \mathcal{N} \sim 50 - 60 \). Here $\phi_e$ is the field value at the end of reheating and its contribution to Eq.~\eqref{N_e} is negligible since $\phi_*\gg \phi_e$. Hence, we have a simple relation
\begin{align}
 \phi_*\simeq \frac{1}{a}\ln \left(a^2 c_n \mathcal{N}\right), \quad {\rm{or}}  \quad \tau_2^*\simeq a^2 c_n \mathcal{N}.  
\end{align}

The important observation here is that the number of species $N$ or the species scale $\Lambda_{\rm{sp}}$ is related to the field value $\tau_2^*$ during inflation through Eqs.~\eqref{N_sp} and~\eqref{eq:L_sp} in large field limit $\tau_2^*\gg 1$. Therefore, one can directly relate the number of e-folds $\mathcal{N}$ and the species scale $\Lambda_{\rm sp}$ by
\begin{align}
    {\cal N} \simeq \frac{\tau_2^\ast}{a^2 c_n} \simeq \frac{N}{2\pi a^2 c_n }\simeq \frac{1}{2\pi a^2 c_n }\left( \frac{M_{\rm Pl}}{\Lambda_{\rm sp}}\right)^2,\label{N_e_L}
\end{align}
which allows us to express inflationary observable by~$\Lambda_{\rm sp}$. 

Let us apply Eq.~\eqref{N_e_L} to the spectral index~$n_s=1-6 \epsilon+2 \eta$ and the tensor-to-scalar ratio $r=16 \epsilon$. Using Eq.~\eqref{N_e_L}, they can be estimated as
\begin{align}
&n_s\simeq 1-\frac{2}{\mathcal{N}}\simeq 1-4\pi a^2c_n\left(\frac{\Lambda_{\rm{sp}}}{\Mpl}\right)^2,\\
&r \simeq \frac{8}{a^2 \mathcal{N}^2}\simeq 32\pi^2a^2c_n^2\left(\frac{\Lambda_{\rm{sp}}}{\Mpl}\right)^4.
\label{eq:r}
\end{align}
Thus, by comparing the above expressions to observations~\cite{Planck:2018jri}, we can obtain constraints on (or determine the value of) the species scale. For example, from the best-fit value of the spectral index \( n_s = 0.9649 \), we can fix the species scale as
\begin{align}
\Lambda_{\mathrm{sp}} \simeq  8.97 \times 10^{16} \left( \frac{1}{a} \right) \left( \frac{2}{c_n} \right)^{1/2} \, (\mathrm{GeV}), \label{L_n_s}
\end{align}
where we set the reduced Planck
 scale by \( \Mpl = 2.4 \times 10^{18} \, (\mathrm{GeV}) \). Additionally, from the constraint on the tensor-to-scalar ratio \( r < 0.035 \) \cite{Planck:2018jri}, we obtain \( \Lambda_{\rm{sp}} \lesssim 1.74 \times 10^{17} \, (\mathrm{GeV}) \) for \( a = 1 \) and \( c_n = 2 \), which is automatically satisfied by Eq.~\eqref{L_n_s}. The result in Eq.~\eqref{L_n_s} is quite suggestive, as the modular inflation models combined with CMB observations indicate the presence of a new scale around the scale of Grand Unification Theory 
(GUT), \( \sim 10^{16} \, (\mathrm{GeV}) \). 

Note that the overall size of the scalar potential \( V_0 \) is fixed by \( V_0 \simeq 2.5 \times 10^{-7} \Mpl^4 / (a^2 \mathcal{N}^2) \) to satisfy the CMB normalization on the scalar power spectrum: \( A_s \equiv V / (24 \pi^2 \epsilon) \simeq V_0 a^2 \mathcal{N}^2 / (12 \pi^2) \), which matches the observed value \( A_s \simeq 2.1 \times 10^{-9} \). The Hubble scale during inflation can then be estimated as \( H_{\rm{inf}} \simeq \sqrt{V_0 / (3 \Mpl^2)} \simeq 6.91 \times 10^{14} / (a \mathcal{N}) \, (\mathrm{GeV}) \). This gives \( H_{\rm{inf}} \simeq 1.15 \times 10^{13} \, (\mathrm{GeV}) \) for \( \mathcal{N} = 60 \) and \( a = 1 \), which satisfies \( H_{\rm{inf}} < \Lambda_{\rm sp} \), ensuring that our models remain within a valid EFT description.

\section{Generic Calabi-Yau cases}
Let us discuss the more generic case. Thus far, we have adopted a specific moduli dependence of the species scale \eqref{N_sp}, but the species scale in generic CY threefolds \eqref{eq:N_sp_general} leads to the following relation between the number of e-folding and the species scale:
\begin{align}
    {\cal N} \simeq \frac{12}{2\pi a^2 c_n c_2^{({\rm CY})}} \left( \frac{M_{\rm Pl}}{\Lambda_{\rm sp}}\right)^2,
\label{eq:N_general}
\end{align}
where we assume that the inflaton potential for a generic modulus is still given by Eq.~\eqref{V_inf}. \footnote{This is a non-trivial assumption; however, it still follows that inflationary observables constrain the species scale if the modulus determines both the inflationary dynamics and the value of species scale.} 
In the same way as the previous section, from the slow-roll approximation for the spectral index $n_s\simeq 1-2/{\cal N}$ with Eq.~\eqref{eq:N_general}, the observed value of the spectral index fixes the species scale:
\begin{align}
\Lambda_{\mathrm{sp}} \simeq 4.39 \times 10^{16} \left( \frac{1}{a} \right) \left( \frac{2}{c_n} \right)^{1/2} \left( \frac{50}{c_2^{({\rm CY})}} \right)^{1/2} \, (\mathrm{GeV}). 
\label{eq:L_n_s_general}
\end{align}

\begin{figure}[htb]
    \centering
    \includegraphics[width=1\linewidth]{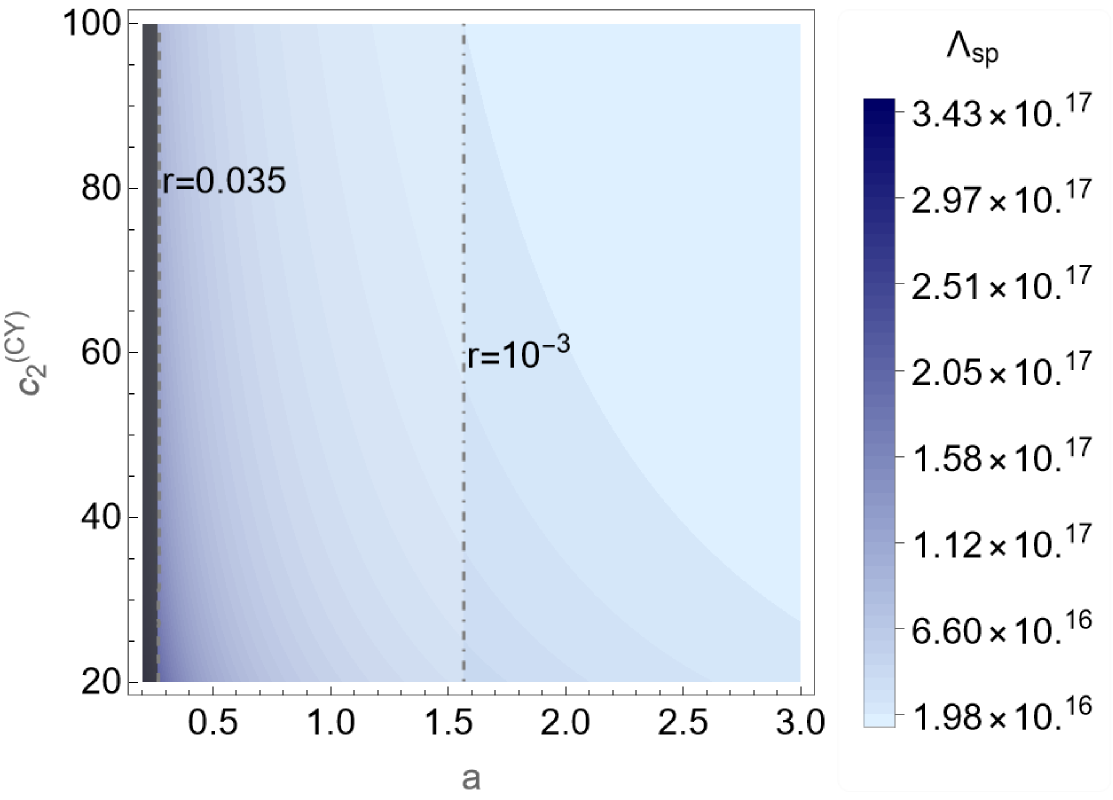}
    \caption{Values of the species scale $\Lambda_{\rm sp}$ in units of GeV as functions of $a$ and $c_2^{({\rm CY})}$ with $c_n=2$. The dashed and dotdashed curves respectively correspond to $r=0.035$ and $r=10^{-3}$ by employing Eq. \eqref{eq:r} for the expression of $r$. The dark gray region with $r\gtrsim 0.035$ is excluded by the Planck data \cite{Planck:2018jri}.}
    \label{fig}
\end{figure}

Note that it is possible to consider a wide range for the second Chern number of CY threefolds. For instance, a typical value of the second Chern numbers in complete intersection CY threefolds is of ${\cal O}(50)$~\cite{CICY}. 
To see the parameter dependence of the species scale, we plot $\Lambda_{\mathrm{sp}}$ as functions of $a$ and $c_2^{({\rm CY})}$ with $c_n=2$ in Fig. \ref{fig}, where the dark gray region with $r\gtrsim 0.035$ is excluded by the current Planck data \cite{Planck:2018jri}. 
Note that the tensor-to-scalar ratio is also written by
\begin{align}
   r \simeq \frac{8}{a^2 \mathcal{N}^2}
   \simeq \frac{2.46\times 10^{-3}}{a^2},
\end{align}
where we use Eqs.~\eqref{eq:N_general} and~\eqref{eq:L_n_s_general} in the last inequality. 
In particular, the region with small \( a \) is excluded in current observations. 
On the other hand, the region with $r\gtrsim 10^{-3}$ will be probed by the next generation of CMB polarization experiments such as CMB-S4 \cite{CMB-S4:2020lpa} and LiteBIRD \cite{LiteBIRD:2022cnt}, and the detection of primordial gravitational waves will lead to the determination of the species scale for each inflationary model, irrelevant to the value of the second Chern numbers of the underlying CY threefolds.

Finally, we comment on the scale dependence of \( n_s \), known as the running of the spectral index \( \alpha_s \) and its running \( \beta_s \), which can be expressed in terms of the slow-roll parameters, including higher derivatives of the scalar potential~\cite{Zarei:2014bta}, and subsequently fixed by the spectral index:
\begin{align}
    \alpha_s &= 16\epsilon \eta - 24\epsilon^2 - 2\xi \simeq -2 \eta^2 \simeq -8.0 \times 10^{-3} \left(\frac{n_s}{0.96} \right)^2,
    \nonumber\\
    \beta_s &= 2\eta \xi + 2\sigma - 24\epsilon \xi -32\epsilon \eta^2 + 192\epsilon^2 \eta - 192 \epsilon^3 
    \simeq 4\eta^3
    \nonumber\\
    &\simeq -3.2\times 10^{-5} \left(\frac{n_s}{0.96} \right)^3,
\end{align}
with slow-roll parameters:
\begin{align}
&\xi \equiv \Mpl^4\frac{V' V'''}{V^2}\simeq  \eta^2,
\qquad
\sigma \equiv \Mpl^6\frac{(V')^2 V''''}{V^3}\simeq \eta^3,
\nonumber\\
&\chi \equiv \Mpl^8\frac{(V')^3 V'''''}{V^4}\simeq \eta^4.
\end{align}
Hence, by comparing with the observational constraints of future experiments, one can obtain further constraints on the parameter spaces of \( a \), \( c_n \), and \( c_2^{({\rm CY})} \) on top of Eq.~\eqref{eq:L_n_s_general}.

\vspace{-0.1cm}

\section{Conclusion}
\label{sec:conclusions}
In this {\it{Letter}}, we pointed out that the species scale of modular inflation models can be constrained (or almost determined) by confronting the theoretical predictions of inflationary observables with observational constraints. Specifically, we consider $SL(2, \mathbb{Z})$-modular inflation models and express the spectral index and tensor-to-scalar ratio in terms of the species scale. These can be directly constrained by the Planck data, from which we obtain $\Lambda_{\rm sp} \sim 10^{16} \, {\rm GeV}$ for reasonable choices of parameters. 
Here, we discussed a more general setup with generic Calabi-Yau threefolds and found that our conclusion remains robust. Therefore, in a large class of modular inflation models where the moduli act as both the inflaton and the field determining the species scale, we expect a new scale to emerge.

Remarkably, the species scale predicted here is situated around the GUT scale, which is promising. This indicates that gravitational effects could play an important role in GUT physics. Revisiting GUT-scale physics by evaluating higher-dimensional operators induced by gravity, would be an interesting direction for future work. 
In particular, the appearance of a low cutoff scale compared to the Planck scale has important cosmological implications. For example, in the context of multifield inflation, a large turn in a curved field-space trajectory can generate non-Gaussianity in the primordial curvature perturbation. However, realizing such a large turn often requires sub-Planckian physics~\cite{Aoki:2024jha}. Furthermore, the presence of a low cutoff scale characterizing the interactions between the inflaton and massive fields has been discussed in the context of cosmological collider physics with a chemical potential~\cite{Chen:2018xck, Bodas:2020yho}.

\vspace{-0.2cm}

\acknowledgments
\vspace{-0.2cm}
\noindent
S.A. is supported by IBS under the project IBS-R018-D1. This work was supported in part by JSPS KAKENHI Grant Numbers JP23H04512 (H.O).


\bibliographystyle{utphys}
\bibliography{references}

\end{document}